\documentclass[conference]{IEEEtran}

\usepackage{amsmath}
\usepackage{amsfonts}
\usepackage{graphicx}
\usepackage[short]{optidef}
\usepackage{newtxtext,newtxmath} 
\usepackage{mleftright} 
\mleftright 
\usepackage[margin=0.70in]{geometry}
\usepackage{cite}
\newcommand{\Fourier}{\mbox{\boldmath$\cal F$}}

\newcommand {\balpha} {\mbox{\boldmath{$\alpha$}}}
\newcommand {\bbeta} {\mbox{\boldmath{$\beta$}}}
\newcommand {\nth} {$n^{\textrm{th}}\ $}
\newcommand {\mth} {$m^{\textrm{th}}\ $}

\IEEEoverridecommandlockouts
\begin{document}

\title{Frequency Domain Multi-user Detection for Single Carrier Modulation with Cyclic Prefix}
\date{29 May 2020}
 \author{%
    \IEEEauthorblockN{Brent A. Kenney\IEEEauthorrefmark{1}, Arslan J. Majid\IEEEauthorrefmark{2}, Hussein Moradi\IEEEauthorrefmark{2}, and Behrouz Farhang-Boroujeny\IEEEauthorrefmark{1}}
    \IEEEauthorblockA{\IEEEauthorrefmark{1}Electrical and Computer Engineering Department, University of Utah, Salt Lake City, Utah, USA}
    \IEEEauthorblockA{\IEEEauthorrefmark{2}Idaho National Laboratory, Salt Lake City, Utah, USA\\}
\thanks{This manuscript has in part been authored by Battelle Energy Alliance, LLC under Contract No. DE-AC07-05ID14517 with the U.S. Department of Energy. The United States Government retains and the publisher, by accepting the paper for publication, acknowledges that the United States Government retains a nonexclusive, paid-up, irrevocable, world-wide license to publish or reproduce the published form of this manuscript, or allow others to do so, for United States Government purposes. \bf{STI Number: INL/CON-20-58011}.}
}

\maketitle
\begin{abstract}
Frequency domain (FD) multi-user detection (MUD) has been shown to be an effective means of approaching the theoretical per-user capacity for single carrier modulation (SCM) schemes in massive MIMO scenarios with highly dispersive channels.  When a cyclic prefix is added to the SCM waveform, the circulant structures of the resulting convolutional channel matrix allows for relatively simple expressions for the FD detection.  In this paper, we develop a computationally efficient minimum mean squared error (MMSE) FD-MUD technique in a time-division duplexing (TDD) massive MIMO setup and show how processing steps can be shared between uplink and downlink. 
\end{abstract}

\section{Introduction}
Over the past several years, research on massive multiple input multiple output (MIMO) wireless communications has produced several encouraging findings.  By eliminating inter-symbol interference (ISI) through the use of orthogonal frequency division multiplexing (OFDM) modulation, researchers have shown that linear detectors and precoders can achieve the promised massive MIMO effect of spatially isolating signals from different users \cite{Hoydis:2013}.  In particular, the results in \cite{Hoydis:2013} show that even matched filter (MF) detection is a sufficient linear detector and time-reversal (TR) precoding is a sufficient linear precoder, given enough base station (BS) antennas.

While OFDM is the dominant waveform in current wireless standards, it has a number of drawbacks.  Perhaps the most notorious is the high peak-to-average power ratio (PAPR) associated with OFDM, since the high PAPR requires operation with a high power back-off, resulting in low power efficiency \cite{Tse:2005}.  One common solution to the PAPR problem is to employ some form of single-carrier modulation (SCM), particularly in the uplink (UL) direction, where the transmit power efficiency of the user equipment (UE) is of chief importance.  

Switching modulation formats from OFDM to SCM presents other challenges.  Among these challenges is dealing with the ISI of highly dispersive channels.  Many promising equalization schemes have been devised and summarized in \cite{Benvenuto:2010} for dealing with ISI for SCM waveforms.  It should be noted that the performance gains achieved in \cite{Benvenuto:2010} also come with the cost of a cyclic prefix (CP), similar to OFDM.  However, the methods in \cite{Benvenuto:2010} do not focus on the multi-user scenario with a multi-antenna BS that is targeted by massive MIMO systems.

Earlier work reported in \cite{Larsson:2012} showed that with simple TR precoding for the downlink (DL), SCM operation was nearly optimal for massive MIMO in the low signal-to-noise ratio (SNR) regime.  Similar results are obtained by using MF detection with maximal ratio combining (MRC) (a.k.a. TR-MRC) for the UL \cite{Ngo:2013}, but the ISI and multi-user interference (MUI) dominate when operating at moderate-to-high SNR values.

The recent work in \cite{Farhang:2020} addressed both the ISI and the MUI impacts in the UL by using a minimum mean squared error (MMSE) linear multi-user detector (MUD) on the intermediate results obtained through TR-MRC detection in the frequency domain (FD). The authors compared their FD-MUD technique with the time domain (TD) algorithm proposed in \cite{Torlak:2018} (i.e., SC-ZF-LE).  The complexity of the SC-ZF-LE approach varied not only by the number of users but also by the length of the channel and equalizer, resulting in a much higher computational cost.  Despite the higher complexity, one advantage of the SC-ZF-LE is that performance is nearly optimal even at high input SNR values, which was a shortcoming of the FD-MUD in \cite{Farhang:2020}.

The authors in \cite{Farhang:2020} addressed a continuous SCM signal, processed in blocks, resulting in some approximations that led to sub-optimal performance at higher SNR values.  In addition, the high-SNR performance was heavily dependent on the block size used in the FD processing.  Interestingly, these approximations can be entirely avoided by adding a CP to each frame of transmit data, resulting in nearly optimal performance even in the high SNR regime.  This paper shows how the FD-MUD is adapted to CP-based multi-user processing, hereafter referred to as MRC-MMSE.  One of the main contributions made in this paper is to show that the MRC-MMSE detector is mathematically equivalent to the conventional MMSE detector and leads to a significant computational complexity saving in massive MIMO systems.

The work reported in \cite{Dinis:2018} analyzed a FD precoding scheme based on zero-forcing (ZF) equalization that was developed for the DL in a CP-aided SCM massive MIMO setup. Our results in this paper show that the complexity of the UL MRC-MMSE detector is comparable with the DL FD-ZF precoder.  Moreover, by sharing the matrix inversion performed for the UL with the DL precoder, we introduce a UL/DL transceiver structure with minimal additional computational complexity. 


\section{System Model} \label{System_Model}
The scenario modeled in this paper assumes that the user equipment (UE) does not have any channel state information (CSI).  Each UE is assumed to have knowledge of the average channel power so that it can maintain a power target at the BS, averaged across all BS antennas.  The BS has CSI between each UE and each of the $M$ BS antennas, which can be obtained by transmitting pilot signals from the UEs.  This paper assumes perfect CSI at the BS and leaves the analysis of channel estimation error for a future work.

The uplink (UL) transmission is divided into frames, where each user transmits $N$ unit variance symbols, represented as $\mathbf{s}_k$ for the $k^{\textrm{th}}$ UE.  A cyclic prefix (CP) is added to the front of the frame to preserve circular convolution.  The received signal at the $m^{\textrm{th}}$ antenna after CP removal is expressed  as
\begin{equation}
    \mathbf{y}_m = \sum_{ k=1}^K \mathbf{H}_{m,k} \mathbf{s}_k  + \mathbf{w}_m, 
    \label{eq:01}
\end{equation}
where $\mathbf{H}_{m,k}$ is the $N \times N$ convolutional channel matrix between antenna $m$ and user $k$, and $\mathbf{w}_m$ is the receiver noise ($\mathbf{w}_m \sim \mathcal{N}( \mathbf{0}, \sigma^2_w \mathbf{I}_N )$, where $\mathbf{I}_N$ is the identity matrix of dimension $N$).  Let $\mathbf{h}_{m,k}$ represent the channel impulse response vector between antenna $m$ and user $k$, which is of length $L_h$.  $\mathbf{H}_{m,k}$ is formed by taking $\mathbf{h}_{m,k}$, appending $N-L_h$ zeros to form ${\mathbf{h}_{m,k} }_{(0)}$, and then taking downward cyclic shifts of ${ \mathbf{h}_{m,k} }_{(0)}$ to create
\begin{equation}
  \mathbf{H}_{m,k}=
  \begingroup 
    \setlength\arraycolsep{4pt}
    \begin{bmatrix}
       { \mathbf{h}_{m,k} }_{(0)} & { \mathbf{h}_{m,k} }_{(1)} \dots & { \mathbf{h}_{m,k} }_{(N-2)} & { \mathbf{h}_{m,k} }_{(N-1)}
    \end{bmatrix}
  \endgroup, 
    \label{eq:02}
\end{equation}
where the parenthetical subscript represents the number of cyclical shifts applied to the base vector.  

For convenience and without loss of generality, the average channel power for each user over all BS antennas, $\frac{1} {M} \sum_{m=1}^M \mathbf{h}_{m,k}^{\textrm{H}} \mathbf{h}_{m,k}$, is normalized to unity.  Consequently, the input SNR is ${ 1 }/ {\sigma^2_w}$.  Note that the individual channel power values between antenna $m$ and user $k$ can vary widely while still maintaining the target average power.

\section{ Frequency Domain Multi-user Detection} \label{UL_MRC-MMSE}
We now introduce two versions of FD-MUD for SCM signals in the UL.  The advantage of performing detection in the FD is that the received signal for each user is equalized (i.e., ISI is removed) in addition to cancelling the MUI.  The first approach is a straightforward application of a MMSE detector applied in the FD.  The MRC-MMSE detector is the second approach, which generates an intermediate result through MF and MRC and then applies the MMSE detector.

In order to process the received signal in the FD, we take the $N$-point Discrete Fourier Transform (DFT) of the received signal vector in \eqref{eq:01}, which results in the expression
\begin{equation}
    \tilde{ \mathbf{y} }_m = \sum_{k=1}^K \mathbf{\Lambda}_{m,k} \tilde{ \mathbf{s} }_k + \tilde{ \mathbf{w} }_m, 
    \label{eq:04}
\end{equation}
where the tilde represents the FD representation of the vectors.  The matrix $\mathbf{\Lambda}_{m,k}$ results from the fact that the circulant matrix $\mathbf{H}_{m,k}$ is diagonalized by the DFT matrix,  $\Fourier$, where $\Fourier$ is scaled such that $\Fourier^{-1} = \Fourier^{\textrm{H}}$ (i.e., $\mathbf{H}_{m,k} = \Fourier^{-1} \mathbf{\Lambda}_{m,k} \Fourier$) \cite{Farhang:Adaptive_Filters}.  It follows that taking the $N$-point DFT results in $\Fourier \mathbf{H}_{m,k} = \mathbf{\Lambda}_{m,k} \Fourier$, where $\mathbf{\Lambda}_{m,k}$ is a diagonal matrix containing the eigenvalues of $\mathbf{H}_{m,k}$.  Let $\lambda_{m,k,i}$ represent the $i^{\textrm{th}}$ value along the diagonal of $\mathbf{\Lambda}_{m,k}$.  The eigenvalues can also be obtained by taking the N-point DFT of the channel impulse response $\mathbf{h}_{m,k}$, which is used to form $\mathbf{H}_{m,k}$.  For more efficient computation, it is noted that all of the FD conversions can be performed with the Fast Fourier Transform (FFT) instead of the DFT.

\subsection{MMSE Detector}
The MMSE detector is calculated in the FD for each bin.  We represent the $n^{\textrm{th}}$ bin of the FD vectors in \eqref{eq:04} with an additional subscript.  The \nth bin of the $M$ input signals can be organized together in vector form as follows:
\begin{equation} \small
	\begin{bmatrix}
    	  \tilde{y}_{1,n} \\
	  \tilde{y}_{2,n} \\
	  \vdots \\
	  \tilde{y}_{M,n}
	\end{bmatrix} =
	\begin{bmatrix}
    	  \lambda_{1,1,n} & \lambda_{1,2,n} & \dots & \lambda_{1,K,n} \  \\
    	  \lambda_{2,1,n} & \lambda_{2,2,n} & \dots & \lambda_{2,K,n} \  \\
	  \vdots \\
    	  \lambda_{M,1,n} & \lambda_{M,2,n} & \dots & \lambda_{M,K,n} \  \\
	\end{bmatrix} 
	\begin{bmatrix}
    	  \tilde{s}_{1,n} \\
	  \tilde{s}_{2,n} \\
	  \vdots \\
	  \tilde{s}_{K,n}
	\end{bmatrix} +
	\begin{bmatrix}
    	  \tilde{w}_{1,n} \\
	  \tilde{w}_{2,n} \\
	  \vdots \\
	  \tilde{w}_{M,n}
	\end{bmatrix}. 
    \label{eq:05}
\end{equation}
For simpler notation of \eqref{eq:05}, the FD vector of received values will be referred to as $\tilde{ \mathbf{y} }_{:,n}$, the matrix of eigenvalues will be represented as $\mathbf{A}_n$, the FD vector of transmitted symbols will be referred to as $\tilde{ \mathbf{s} }_{:,n}$, and the FD vector of noise samples will be referred to as $\tilde{ \mathbf{w} }_{:,n}$.  The simplified expression for \eqref{eq:05} is now expressed as
\begin{equation}
    \tilde{ \mathbf{y} }_{:,n} = \mathbf{A}_n \tilde{ \mathbf{s} }_{:,n} + \tilde{ \mathbf{w} }_{:,n}. 
    \label{eq:06}
\end{equation}

The MMSE solution to \eqref{eq:06} from Theorem 10.3 of \cite{Kay:1993} is given as
\begin{equation}
    \hat{\tilde{\mathbf{s}}}_{:,n}^{\textrm{MMSE}}= \mathbf{a} \circ \mathbf{A}_n^{\textrm{H}} \left(  \mathbf{A}_n \mathbf{A}_n^{\textrm{H}} + \sigma_w^2 \mathbf{I}_M \right)^{-1} \tilde{\mathbf{y}}_{:,n}, 
    \label{eq:07}
\end{equation}
where $\mathbf{a}$ is included as a vector of coefficients to ensure that the estimates are unbiased, $\circ$ is the Haddamard multiply operator, $\sigma_w^2 \mathbf{I}_M$ is the covariance of the FD noise vector, and $(\cdot )^{\textrm{H}}$ is the Hermitian operator.  By combining \eqref{eq:06} and \eqref{eq:07}, we calculate the elements of $\mathbf{ a }$ to be the inverse of the diagonal values of $\mathbf{A}_n^{\textrm{H}} \left(  \mathbf{A}_n \mathbf{A}_n^{\textrm{H}} + \sigma_w^2 \mathbf{I}_M \right)^{-1} \mathbf{A}_n$ in order for the estimate to be unbiased.  This can also be expressed as
\begin{equation}
    \mathbf{a} = \textrm{inv} \left( \textrm{diag} \left( \mathbf{A}_n^{\textrm{H}} \left(  \mathbf{A}_n \mathbf{A}_n^{\textrm{H}} + \sigma_w^2 \mathbf{I}_M \right)^{-1} \mathbf{A}_n \right) \right), \label{eq:07a}
\end{equation}
where the $\textrm{inv}(\cdot)$ operator performs an element-wise inverse, and the $\textrm{diag}(\cdot)$ operator arranges the diagonal elements of the $M \times M$ matrix operand in a $M \times 1$ vector such that the $i^{\textrm{th}}$ element of the vector result is equal to the element from the $i^{\textrm{th}}$ row and $i^{\textrm{th}}$ column of the matrix operand.

After \eqref{eq:07a} and \eqref{eq:07} are computed for each bin, the $K \times 1$ vector of FD symbol estimates from each of the $N$ bin computations are organized into a $N \times 1$ vector for each of the users.  The resulting estimates are converted back to the TD through an inverse Fast Fourier Transform (IFFT).

\subsection{MRC-MMSE Detector}
The MRC-MMSE detector takes the $M$ received vectors in \eqref{eq:01} and calculates intermediate values through MF and MRC.  This method was used in \cite{Farhang:2020} for SCM without a CP, which resulted in extra noise and interference terms that required some approximation to form tractable expressions.  By adding the CP, no approximation to circular convolution is needed, and the additional interference terms, which resulted in varying levels of performance depending on frame size, are no longer applicable.  

The MF operation is performed with a time-reversed (TR) and conjugated version of the channel impulse response, which is equivalent to multiplying the \mth received vector by the Hermitian of the channel matrix corresponding to the \mth antenna and user $k$.  Since the MF operation emphasizes signals with higher channel powers and de-emphasizes signals with lower channel powers, averaging the results from all of the $M$ antennas is MRC.  Consequently, the intermediate results are called TR-MRC estimates, as used in \cite{Farhang:2020}.  We represent the TR-MRC value for each UE as
\begin{align}
    \mathbf{r}_k &= \frac{1}{M} \sum_{m=1}^M \mathbf{H}_{m,k}^{\textrm{H}} \mathbf{y}_m.
    \label{eq:08}
\end{align}

Although the computation of the TR-MRC values are shown in the TD, this process is easily incorporated into the FD processing.  We make use of the diagonalization of the circulant matrix $\mathbf{H}_{m,k}^{\textrm{H}}$, and represent it in the FD as $\Fourier \mathbf{H}_{m,k}^{\textrm{H}} =  \mathbf{\Lambda}_{m,k}^* \Fourier$, where $(\cdot )^*$ is the conjugate operator.  The FD representation of \eqref{eq:08} is now given as
\begin{align}
	\tilde{\mathbf{r}}_k &= \frac{1}{M} \sum_{m=1}^M  \mathbf{\Lambda}_{m,k}^* \tilde{ \mathbf{ y } }_m.
    \label{eq:09}
\end{align}

We observe that the $n^{\textrm{th}}$ element of $\tilde{\mathbf{r}}_k$ is a weighted summation of the $M$ receive signals at the $n^{\textrm{th}}$ frequency bin.  We now express the TR-MRC result for each user on a frequency bin basis.  
\begin{equation} \small
	\begin{bmatrix}
    	  \tilde{r}_{1,n} \\
	  \tilde{r}_{2,n} \\
	  \vdots \\
	  \tilde{r}_{K,n}
	\end{bmatrix} = \frac{1}{M}
	\begin{bmatrix}
    	  \lambda_{1,1,n}^* & \lambda_{2,1,n}^* & \dots & \lambda_{M,1,n}^* \  \\
    	  \lambda_{1,2,n}^* & \lambda_{2,2,n}^* & \dots & \lambda_{M,2,n}^* \  \\
	  \vdots \\
    	  \lambda_{1,K,n}^* & \lambda_{1,K,n}^* & \dots & \lambda_{M,K,n}^* \  \\
	\end{bmatrix} 
	\begin{bmatrix}
    	  \tilde{y}_{1,n} \\
	  \tilde{y}_{2,n} \\
	  \vdots \\
	  \tilde{y}_{M,n}
	\end{bmatrix}. 
    \label{eq:10}
\end{equation}

The matrix of eigenvalues defined in \eqref{eq:10} is equal to the Hermitian of the $\mathbf{A}_n$ matrix from \eqref{eq:05} and \eqref{eq:06}.  Consequently, the TR-MRC result can be represented compactly as
\begin{align}
	\tilde{\mathbf{r}}_{:,n} &= \frac{1}{M} \mathbf{A}_n^{\textrm{H}} \tilde{\mathbf{y}}_{:,n} \nonumber \\
	&= \frac{1}{M} \mathbf{A}_n^{\textrm{H}} \mathbf{A}_n \tilde{ \mathbf{s} }_{:,n} + \frac{1}{M} \mathbf{A}_n^{\textrm{H}} \tilde{ \mathbf{w} }_{:,n}, \label{eq:11}
\end{align}
where the second expression results from substituting \eqref{eq:06}.  In order to compute the MMSE estimate for $\tilde{\mathbf{s}}_{:,n}$ with the TR-MRC inputs, the noise covariance is required, which is computed as follows:
\begin{align}
	\mathbb{E}\left[ \left( \frac{1}{M} \mathbf{A}_n^{\textrm{H}} \tilde{ \mathbf{w} }_{:,n} \right)  \left( \frac{1}{M} \mathbf{A}_n^{\textrm{H}} \tilde{ \mathbf{w} }_{:,n} \right)^{\textrm{H}} \right] &= \frac{1}{M^2} \mathbb{E} \left[ \mathbf{A}_n^{\textrm{H}}  \tilde{ \mathbf{w} }_{:,n} \tilde{ \mathbf{w} }_{:,n}^{\textrm{H}} \mathbf{A}_n \right] \nonumber \\
	&=\frac{1}{M^2} \mathbf{A}_n^{\textrm{H}}  \left( \sigma^2_w \mathbf{I}_M \right) \mathbf{A}_n \nonumber \\
	&=\frac{\sigma^2_w}{M^2} \mathbf{A}_n^{\textrm{H}} \mathbf{A}_n . \label{eq:12}
\end{align}

Using this result for the covariance, the MRC-MMSE solution can now be expressed per Theorem 10.3 of \cite{Kay:1993} as
\begin{align}
	\hat{\tilde{\mathbf{s}}}_{:,n}^{\textrm{MRC-MMSE}} &= \balpha \circ \frac{ \mathbf{A}_n^{\textrm{H}} \mathbf{A}_n }{ M }  \left(  \frac{1}{M^2} \mathbf{A}_n^{\textrm{H}} \mathbf{A}_n  \mathbf{A}_n^{\textrm{H}} \mathbf{A}_n  +\frac{\sigma^2_w}{M^2} \mathbf{A}_n^{\textrm{H}} \mathbf{A}_n \right)^{-1} \tilde{\mathbf{r}}_{:,n} \nonumber \\
	&= \balpha \circ  M \mathbf{A}_n^{\textrm{H}} \mathbf{A}_n  \left(  \left(  \mathbf{A}_n^{\textrm{H}} \mathbf{A}_n  +\sigma^2_w \mathbf{I}_K \right) \mathbf{A}_n^{\textrm{H}} \mathbf{A}_n \right)^{-1} \tilde{\mathbf{r}}_{:,n} \nonumber \\
	&= \balpha \circ M \left(  \mathbf{A}_n^{\textrm{H}} \mathbf{A}_n  +\sigma^2_w \mathbf{I}_K \right)^{-1} \tilde{\mathbf{r}}_{:,n} \nonumber \\
	&= \balpha \circ \left(  \mathbf{A}_n^{\textrm{H}} \mathbf{A}_n  +\sigma^2_w \mathbf{I}_K \right)^{-1} \mathbf{A}_n^{\textrm{H}} \tilde{\mathbf{y}}_{:,n},
	\label{eq:13}
\end{align}
where $\balpha$  is included as a vector of coefficients to ensure that the estimates are unbiased, and the last expression results from substituting \eqref{eq:11} for $\tilde{\mathbf{r}}_{:,n}$.  Based on the definition of $\tilde{\mathbf{y}}_{:,n}$ in \eqref{eq:06} and the result from \eqref{eq:13}, $\balpha$ is calculated to be 
\begin{equation}
    \balpha = \textrm{inv} \left( \textrm{diag} \left( \left(  \mathbf{A}_n^{\textrm{H}} \mathbf{A}_n  +\sigma^2_w \mathbf{I}_K \right)^{-1} \mathbf{A}_n^{\textrm{H}} \mathbf{A}_n \right) \right). \label{eq:13a}
\end{equation}

The final expression for the MRC-MMSE detector in \eqref{eq:13} has the advantage over the MMSE detector in terms of computational complexity because the MRC-MMSE detector only requires a $K \times K$ matrix inversion per frequency bin instead of a $M \times M$ matrix inversion. After the estimates are computed in the FD for each bin, they are converted back to the TD in the same manner that was described for the MMSE detector.

\subsubsection{MRC-MMSE and MMSE Equivalence}

Prior to analyzing the performance of the MMSE-based detectors, it is important to point out that the MRC-MMSE is a reformulation of the MMSE detector.  Specifically, we prove that
\begin{align}
	\hat{\tilde{\mathbf{s}}}_{:,n}^{\textrm{MMSE}} &= \mathbf{ a } \circ \mathbf{A}_n^{\textrm{H}} \left(  \mathbf{A}_n \mathbf{A}_n^{\textrm{H}} + \sigma^2_w \mathbf{I}_M \right)^{-1} \tilde{\mathbf{y}}_{:,n} \nonumber \\
	&= \balpha \circ \left( \mathbf{A}_n^{\textrm{H}} \mathbf{A}_n + \sigma^2_w \mathbf{I}_K \right)^{-1} \mathbf{A}_n^{\textrm{H}} \tilde{\mathbf{y}}_{:,n} = \hat{ \tilde{ \mathbf{s} } }_{:,n}^{\textrm{MRC-MMSE} }. 
	\label{eq:14}
\end{align}

To this end, we start by specifying the general form of the Matrix Inversion Lemma (MIL) \cite{Woodbury:1950} as
\begin{equation}
	( \mathbf{A} + \mathbf{USV} )^{-1} = \mathbf{A}^{-1} - \mathbf{A}^{-1} \mathbf{U} ( \mathbf{S}^{-1} + \mathbf{VA}^{-1} \mathbf{U})^{-1} \mathbf{VA}^{-1},
	\label{MIL}
\end{equation}
which will be applied to the matrix inversion of the MMSE detector.  We set $\mathbf{A} = \sigma^2_w \mathbf{I}_M$, $\mathbf{U} = \mathbf{A}_n$, $\mathbf{S} = \mathbf{I}_K$, and $\mathbf{V} = \mathbf{A}_n^{\textrm{H}}$.  The result from the MIL substitution is
\begin{equation}
( \sigma^2_w \mathbf{I}_M + \mathbf{A}_n \mathbf{A}_n^{\textrm{H}} )^{-1} = \frac{ \mathbf{I}_M }{\sigma^2_w} - \frac{ \mathbf{I}_M }{\sigma^2_w} \mathbf{A}_n ( \mathbf{I}_K + \mathbf{A}_n^{\textrm{H}} \frac{ \mathbf{I}_M }{\sigma^2_w} \mathbf{A}_n)^{-1} \mathbf{A}_n^{\textrm{H}} \frac{\mathbf{I}_M}{\sigma^2_w}, \label{eq:16}
\end{equation}
since $\left( \sigma^2_w \mathbf{I}_M \right)^{-1} =  \mathbf{I}_M / \sigma^2_w$.  Substituting the above result into \eqref{eq:07} leads to
\begin{align}
	\hat{\tilde{\mathbf{s}}}_{:,n}^{\textrm{MMSE}} &= \mathbf{ a } \circ \mathbf{A}_n^{\textrm{H}} \left( \frac{ \mathbf{I}_M }{\sigma^2_w} - \frac{ \mathbf{I}_M }{\sigma^2_w} \mathbf{A}_n \left( \mathbf{I}_K + \mathbf{A}_n^{\textrm{H}} \frac{ \mathbf{I}_M }{\sigma^2_w} \mathbf{A}_n \right)^{-1} \mathbf{A}_n^{\textrm{H}} \frac{\mathbf{I}_M}{\sigma^2_w} \right) \tilde{\mathbf{y}}_{:,n} \nonumber \\
	&=  \mathbf{ a } \circ \frac{ \mathbf{A}_n^{\textrm{H}} }{\sigma^2_w}  \left( \mathbf{I}_M - \mathbf{A}_n \left( \sigma^2_w \mathbf{I}_K + \mathbf{A}_n^{\textrm{H}} \mathbf{A}_n \right)^{-1} \mathbf{A}_n^{\textrm{H}} \right) \tilde{\mathbf{y}}_{:,n} \nonumber \\
	&=   \mathbf{ a } \circ  \frac{ 1 }{\sigma^2_w} \left( \mathbf{A}_n^{\textrm{H}} - \mathbf{A}_n^{\textrm{H}} \mathbf{A}_n \left( \sigma^2_w \mathbf{I}_K + \mathbf{A}_n^{\textrm{H}} \mathbf{A}_n \right)^{-1} \mathbf{A}_n^{\textrm{H}} \right) \tilde{\mathbf{y}}_{:,n} \nonumber \\
	&=  \mathbf{ a } \circ   \frac{ 1 }{\sigma^2_w} \left( \mathbf{I}_K - \mathbf{A}_n^{\textrm{H}} \mathbf{A}_n \left( \sigma^2_w \mathbf{I}_K + \mathbf{A}_n^{\textrm{H}} \mathbf{A}_n \right)^{-1} \right) \mathbf{A}_n^{\textrm{H}} \tilde{\mathbf{y}}_{:,n} \nonumber \\
	&=  \mathbf{ a } \circ \left( \frac{ \left( \sigma^2_w \mathbf{I}_K + \mathbf{A}_n^{\textrm{H}} \mathbf{A}_n \right) - \mathbf{A}_n^{\textrm{H}} \mathbf{A}_n }{ \sigma^2_w } \right) \left( \sigma^2_w \mathbf{I}_K + \mathbf{A}_n^{\textrm{H}} \mathbf{A}_n \right)^{-1} \mathbf{A}_n^{\textrm{H}} \tilde{\mathbf{y}}_{:,n} \nonumber \\
	&=  \mathbf{ a } \circ  \left( \frac{ \sigma^2_w \mathbf{I}_K }{ \sigma^2_w } \right) \left( \sigma^2_w \mathbf{I}_K + \mathbf{A}_n^{\textrm{H}} \mathbf{A}_n \right)^{-1} \mathbf{A}_n^{\textrm{H}} \tilde{\mathbf{y}}_{:,n} \nonumber \\
	&=  \mathbf{ a } \circ  \left( \mathbf{A}_n^{\textrm{H}} \mathbf{A}_n + \sigma^2_w \mathbf{I}_K \right)^{-1} \mathbf{A}_n^{\textrm{H}} \tilde{\mathbf{y}}_{:,n}. \label{eq:17}
\end{align}

Up to this point, we have shown that
\begin{equation}
    \mathbf{A}_n^{\textrm{H}} \left(  \mathbf{A}_n \mathbf{A}_n^{\textrm{H}} + \sigma^2_w \mathbf{I}_M \right)^{-1} = \left( \mathbf{A}_n^{\textrm{H}} \mathbf{A}_n + \sigma^2_w \mathbf{I}_K \right)^{-1} \mathbf{A}_n^{\textrm{H}}, \nonumber
\end{equation}
and by extension 
\begin{equation}
    \mathbf{A}_n^{\textrm{H}} \left(  \mathbf{A}_n \mathbf{A}_n^{\textrm{H}} + \sigma^2_w \mathbf{I}_M \right)^{-1} \mathbf{A}_n = \left( \mathbf{A}_n^{\textrm{H}} \mathbf{A}_n + \sigma^2_w \mathbf{I}_K \right)^{-1} \mathbf{A}_n^{\textrm{H}} \mathbf{A}_n. \label{eq:17a}
\end{equation}
By using \eqref{eq:17a} with the definition of $\mathbf{a}$ in \eqref{eq:07a} and $\balpha$ in \eqref{eq:13a}, we conclude that $\balpha = \mathbf{a}$.  As a result, the final result in \eqref{eq:17} can be written as $\balpha \circ \left( \mathbf{A}_n^{\textrm{H}} \mathbf{A}_n + \sigma^2_w \mathbf{I}_K \right)^{-1} \mathbf{A}_n^{\textrm{H}} \tilde{\mathbf{y}}_{:,n}$, which is equal to $\hat{\tilde{\mathbf{s}}}_{:,n}^{\textrm{MRC-MMSE}}$ in \eqref{eq:13}.  This concludes our proof.

\subsubsection{Sufficient Statistics}

By showing that the MRC-MMSE detector is identical to the MMSE detector, we conclude that the MRC-MMSE detector is also optimal in the mean square sense.  We can also conclude that no information is lost by taking the $M$ inputs from the BS antenna array and constructing $K$ intermediate TR-MRC results.  Hence, the $K$ intermediate MF values constitute a sufficient statistic for the optimum detector.  As a result, a drastic reduction in computational complexity is available without sacrificing performance when $M \gg K$.


\section{Complexity Comparison}

\begin{figure}[!t]
\centering
\includegraphics[width=3.6in, clip=true, trim=4cm 8.5cm 4cm 9cm ]{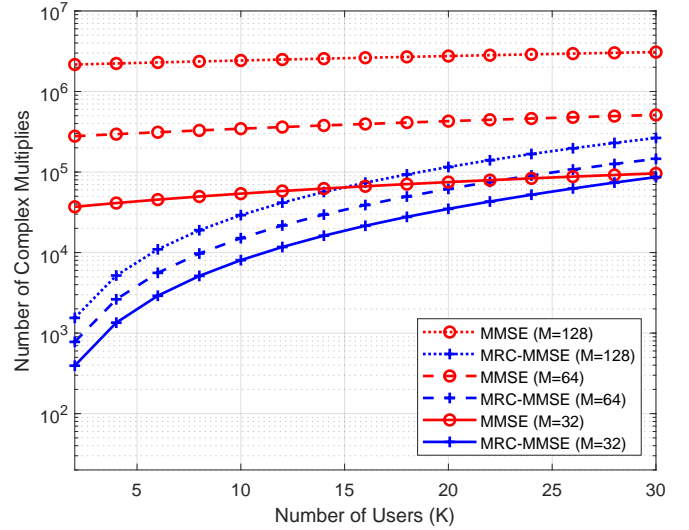}
\caption{The number of complex multiplies per bin are plotted versus the number of users for the two detector implementations and for different numbers of BS antennas (i.e., $M=32$, $64$, and $128$).  The MRC-MMSE approach has lower complexity overall, but especially when $M \gg K$.  }
\label{complexity_comp}
\end{figure}

The difference in the computational complexity between the MMSE and the MRC-MMSE implementations can be seen by comparing the number of complex multiplies.  We can see that the complexity of the MMSE detector far exceeds that of the MRC-MMSE detector when $M \gg K$.

Since both algorithms require conversion to and from the FD, this operation is not considered in the comparison.  In order to evaluate the complexity, we assign $P^3$ complex multiplies for a $P \times P$ matrix inversion.  The number of complex multiples involved in a matrix (or vector) multiplication is the product of the outer dimensions times the inner dimension.  When computing the diagonal elements of the product of two matrices, the number of complex multiplies is the product of the inner dimension and the outer dimension.

The following operations are required for each bin of the MMSE processing from \eqref{eq:07}, resulting in a total of $K + 2KM + 2KM^2 + M^3$ complex multiplies per bin:
\begin{description}
	\item[$KM^2$] Multiply $(M \times K)$ matrix by $(K \times M)$ matrix
	\item[$M^3$] Invert the resulting $(M \times M)$ matrix
	\item[$KM^2$] Multiply $(K \times M)$ matrix by $(M \times M)$ matrix
	\item[$KM$] Multiply $(K \times M)$ matrix by $(M \times 1)$ vector
	\item[$KM$] Calculate diagonals of $(K \times M)$ and $(M \times K)$ product
	\item[$K$] Multiply coefficients by $(K \times 1)$ vector
\end{description}

The last two operations calculate $\mathbf{a}$ and scale the result.  Since $M > K$ in all cases, the $M^3$ term for the MMSE detector quickly begins to dominate the calculation as $M$ grows much larger than $K$.  

The following operations are required for each bin of the MRC-MMSE processing from \eqref{eq:13}, resulting in a total of $K + 2KM + K^3 + 2K^2 M$ complex multiplies per bin:
\begin{description}
	\item[$K^2 M$] Multiply $(K \times M)$ matrix by $(M \times K)$ matrix
	\item[$K^3$] Invert the resulting $(K \times K)$ matrix
	\item[$K^2 M$] Multiply $(K \times K)$ matrix by $(K \times M)$ matrix
	\item[$KM$] Multiply $(K \times M)$ matrix by $(M \times 1)$ vector
	\item[$KM$] Calculate diagonals of $(K \times M)$ by $(M \times K)$ product
	\item[$K$] Multiply coefficients by $(K \times 1)$ vector
\end{description}

Note that the calculation of $\balpha$ for each frequency bin approximately doubles the complex multiplies used in the MRC-MMSE detector.  A method for pre-computing a general scale factor that still produces an unbiased estimate is currently being studied.  Fig. \ref{complexity_comp} shows a comparison of the computational complexity of the MMSE and MRC-MMSE detectors.  As Fig. \ref{complexity_comp} demonstrates, the MRC-MMSE approach has a clear advantage over the MMSE detector for all combinations shown for $M$ and $K$, especially where $M \gg K$.  

\section{Performance Analysis}
In this section, we examine the performance of the MRC-MMSE detector and establish its performance limits for the low and high SNR regimes.  

\subsection{Capacity Limit}
Based on the assumption of uncorrelated channels for each BS antenna and a single-cell setting without pilot contamination, a single-user can see an SNR gain up to a factor of $M$.  If perfect cancellation is achieved, without affecting the array gain, then the average SNR gain in a multi-user scenario would also be scaled by a factor of $M$.  This ideal case can easily be converted to a rate in order to express a capacity by using Shannon's capacity equation
\begin{equation}
	R = \rho W \textrm{log}_2 \left( 1 + \gamma \right), \label{eq:18}
\end{equation}
where $\rho$ is the non-CP duty cycle (e.g., 14/15), $R$ is the achievable bit rate, $W$ is the bandwidth, and $\gamma$ is the effective SNR.  Due to the relationship between the capacity and the effective SNR, we can analyze the performance in terms of the effective SNR.  In particular, we will express the performance gain as the ratio of the average achieved signal to interference-plus-noise ratio (SINR) to the input SNR.

\subsection{Low SNR Operation}
When the noise variance term dominates, the matrix inverse in \eqref{eq:13} and \eqref{eq:13a} will converge to $\frac{1}{\sigma^2_w} \mathbf{I}_K$.  As a result, the scaling term becomes $\balpha = \sigma^2_w \cdot \textrm{inv(diag(} \mathbf{A}_n^{\textrm{H}} \mathbf{A}_n ) )$, and the MRC-MMSE detector simplifies to
\begin{equation}
	\hat{\tilde{\mathbf{s}}}_{:,n}^{\textrm{Low-SNR}} = \textrm{inv} \left( \textrm{diag} \left( \mathbf{A}_n^{\textrm{H}} \mathbf{A}_n \right) \right) \left( \mathbf{A}_n^{\textrm{H}} \mathbf{A}_n \tilde{ \mathbf{s} }_{:,n} + \mathbf{A}_n^{\textrm{H}} \tilde{ \mathbf{w} }_{:,n} \right). \label{eq:19}
\end{equation}
The expression in \eqref{eq:19} can be separated into three parts—the unbiased FD symbol vector, the MUI due to the off-diagonal elements of $\mathbf{A}_n^{\textrm{H}} \mathbf{A}_n$, and the filtered noise.  The MUI component is negligible compared to the noise term in the low SNR regime.  Since $\mathbf{A}_n^{\textrm{H}}$ has entries that are zero-mean, unit-variance, i.i.d. complex Gaussian random variables, the expected value of each term of $\textrm{inv(diag(} \mathbf{A}_n^{\textrm{H}} \mathbf{A}_n ) )$ is $\frac{1}{M}$, and the variance of each element of $\mathbf{A}_n^{\textrm{H}} \tilde{ \mathbf{w} }_{:,n}$ is $M \sigma^2_w$.  Consequently, the variance of the filtered noise vector after scaling by $\frac{1}{M}$ is $\sigma^2_w / M$, which results in an array gain of $M$.

\subsection{High SNR Operation}
At moderate input SNR values, the performance gain diverges from the ideal array gain as the filtered noise term becomes less dominant compared to the MUI.  As the SNR continues to increase into the high SNR regime, the SINR of the detected symbols begins to track a different limit than the ideal array gain.  At high SNR the noise covariance in the inverse term of  \eqref{eq:13} and \eqref{eq:13a} becomes insignificant.  The elements of $\balpha$ converge to unity and the symbol estimates can be expressed as 
\begin{align}
	\hat{\tilde{\mathbf{s}}}_{:,n}^{\textrm{High-SNR}} &= \left(  \mathbf{A}_n^{\textrm{H}} \mathbf{A}_n  \right)^{-1} \mathbf{A}_n^{\textrm{H}} \left( \mathbf{A}_n \tilde{\mathbf{s}}_{:,n} + \tilde{ \mathbf{ w } }_{:,n} \right) \nonumber \\
	&= \tilde{\mathbf{s}}_{:,n}  + \left(  \mathbf{A}_n^{\textrm{H}} \mathbf{A}_n  \right)^{-1} \mathbf{A}_n^{\textrm{H}} \tilde{ \mathbf{ w } }_{:,n}.
	\label{eq:20}
\end{align}

In order to determine the SINR of the estimate at high SNR, we must calculate the scaling associated with the coefficient of the noise, namely $( \mathbf{A}_n^{\textrm{H}} \mathbf{A}_n  )^{-1} \mathbf{A}_n^{\textrm{H}}$.  The expected noise power scaling per user is expressed as
\begin{equation}
    \begin{split}
	\frac{1}{K} \mathbb{E} \left[ \textrm{tr} \left\{ \left( \left( \mathbf{A}_n^{\textrm{H}} \mathbf{A}_n  \right)^{-1} \mathbf{A}_n^{\textrm{H}} \right) \left( \left(  \mathbf{A}_n^{\textrm{H}} \mathbf{A}_n  \right)^{-1} \mathbf{A}_n^{\textrm{H}} \right)^{\textrm{H}} \right\} \right] = \\
	\frac{1}{K} \mathbb{E} \left[ \textrm{tr} \left\{  \left( \mathbf{A}_n^{\textrm{H}} \mathbf{A}_n  \right)^{-1} \right\} \right], 
    \end{split}
\end{equation}
where $\textrm{tr} \left\{ \cdot \right\}$ is the trace operator and the second line results from the fact that $\left( \mathbf{A}_n^{\textrm{H}} \mathbf{A}_n  \right)^{-1}$ is Hermitian symmetric.  From Lemma 6 of \cite{Verdu:2003}, we see that $\mathbb{E} [ \textrm{tr} \{ ( \mathbf{A}_n^{\textrm{H}} \mathbf{A}_n )^{-1} \} ]$ equals $K / (M - K)$.  Hence, the noise seen by each user will be $(M - K)$ times smaller than the input noise, resulting in an array gain of $(M - K)$.

\section{Simulation Results} \label{Simulation_Results}
A single-cell UL scenario is simulated to show the performance of the MRC-MMSE detector.  During each frame, $N=2048$ samples are transmitted.  The $K$ UEs transmit simultaneously to the $M$-element BS without any precoding.  Each channel impulse response between UE $k$ and BS antenna $m$ is randomly selected with an exponential power delay profile and a roll-off factor of 25 samples.  The length of the channel impulse response, $L_h$, is set to 130 samples.  Perfect CSI at the BS is assumed.  

It should be noted that the performance of this algorithm is independent of the value of $L_h$ as long as $L_h \leq L_{CP} - 1$, where $L_{CP}$ is the length of the CP.  It is assumed that each UE has sufficient knowledge of the channel statistics such that the UE can maintain a power control target.  Consequently, the average power of each channel for each user is set to a constant value (unity in this case), even though the individual channel power values are uniformly distributed for each antenna.  In the simulation, the channel power values are uniformly distributed between approximately 0.1 and 1.9.  Slight modifications to the scaling are made to maintain the unity average power constraint after the individual power values are chosen.


The main objective of the simulation is to measure the resulting SINR after the signal has been detected by the multi-user receiver.  The resulting SINR can then be converted to capacity, if desired, using \eqref{eq:18}.  The ideal array gain of $M$ and the array gain of $M-K$ are provided to show asymptotic performance at low and high SNR, respectively.

\begin{figure}[!t]
    \centering
    \includegraphics[width=3.6in, clip=true, trim=4cm 8.5cm 4cm 9cm ]{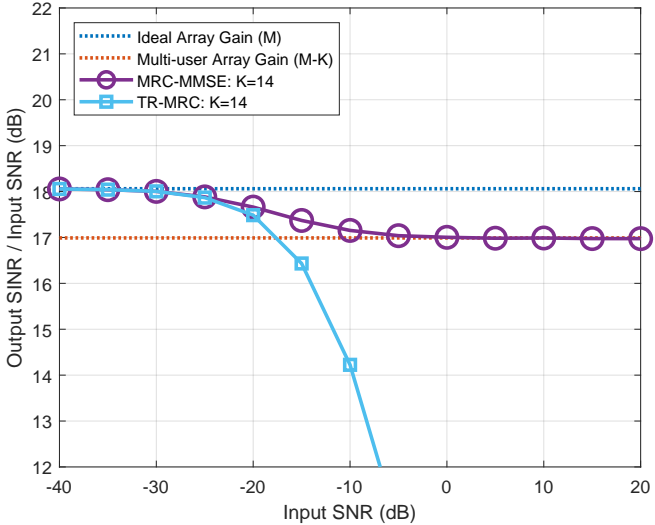}
    \caption{Ratio of output SINR to input SNR is plotted versus the input SNR where $M=64$ and $K=14$.  The MRC-MMSE curve is shown converging to the ideal array gain at low input SNR and to the multi-user array gain of $M-K$ at high input SNR.  The TR-MRC curve is provided for reference.  }
    \label{Input_output_performance_UL}
\end{figure}

The scenario with $M=64$ and $K=14$ is simulated to calculate the output SINR.  In order to better see the transition from the low SNR limit to the high SNR limit, Fig. \ref{Input_output_performance_UL} presents the data with the ordinate selected to be the ratio of the output SINR to the input SNR (i.e., performance gain).  For an example SINR operating point of 10 dB, we see that the input SNR would need to be approximately $-7$ dB.  At this point, the MRC-MMSE performance has largely converged to the $M-K$ performance limit.  The TR-MRC performance is included for comparison.  Even though the MRC-MMSE algorithm is based upon the TR-MRC outputs, TR-MRC alone fails to compensate for ISI and MUI, which start to dominate as the input SNR increases.

\section{Matrix Inversion Resuse for DL}
Once the $K \times K$ matrix inverse in \eqref{eq:13} is computed, it can be reused for DL precoding in a time domain duplex (TDD) system.  The analysis in \cite{Dinis:2018} assumed a zero-forcing precoder for the DL.  However, equivalent performance in the moderate-to-high SNR regime is attained by using the MMSE precoder given by
\begin{equation}
    {\tilde{\mathbf{x}}}_{:,n}^{\textrm{MMSE}} = \bbeta \circ \left(  \mathbf{A}_n^* \mathbf{A}_n^{\textrm{T}} + \mathbf{C}_w \right)^{-1}  \mathbf{A}_n^* \tilde{ \mathbf{s} }_{:,n}, 
    \label{eq:22}
\end{equation}
where ${\tilde{\mathbf{x}}}_{:,n}^{\textrm{MMSE}}$ is the $M \times 1$ precoded vector for bin $n$ in the FD, and $\bbeta$ is a vector of scale factors used to remove bias from the precoded value.  Using the MIL and similar manipulations to those shown in \eqref{eq:17}, we can obtain the following equivalent expression to \eqref{eq:22}:
\begin{equation}
    {\tilde{\mathbf{x}}}_{:,n}^{\textrm{MMSE}} = \bbeta \circ \mathbf{A}_n^* \left(  \mathbf{A}_n^{\textrm{T}} \mathbf{A}_n^* + \sigma^2_w \mathbf{I}_K \right)^{-1}  \tilde{ \mathbf{s} }_{:,n}. 
    \label{eq:23}
\end{equation}

It can be shown that the matrix inverse in \eqref{eq:23} is the complex conjugate of the matrix inverse in \eqref{eq:13}.  This means that the DL precoder can use the same matrix inverse as was calculated for the UL decoder, after conjugating the entries.  Resusing the matrix inversion for the DL reduces the number of precoding computations by an order of magnitude for a moderate number of users (e.g., $K>10$) and even more as $K$ increases.

It should be noted that \eqref{eq:23} assumes the same power is transmitted to all users, irrespective of their distances from the base station.  In order to generalize the precoding result, we introduce a diagonal matrix, $\mathbf{P}^{1/2}$, which contains an amplitude scalar for each user.  This value is set to account for differences in the average path loss to each user.  The precoding equation with generalized amplitude scaling is given by
\begin{equation}
    {\tilde{\mathbf{x}}}_{:,n}^{\textrm{MMSE}} = \bbeta \circ \mathbf{A}_n^* \left(  \mathbf{A}_n^{\textrm{T}} \mathbf{A}_n^* + \sigma^2_w \mathbf{I}_K \right)^{-1}  \mathbf{P}^{1/2} \tilde{ \mathbf{s} }_{:,n}. 
    \label{eq:24}
\end{equation}

Performance of the DL precoding with matrix reuse is currently under study.  Results will be reported in the near future.

\section{Conclusion}
This paper showed that when the FD-MUD detector presented in \cite{Farhang:2020} was adapted to operate with a cyclic prefix, the approximations made in the detector logic are no longer needed.  The resulting theoretical performance of the detector is improved, especially at high input SNR.  We also showed the computational savings of using the MRC-MMSE solution over a conventional MMSE approach.  By proving the mathematical equivalence between the MRC-MMSE and conventional MMSE detectors, we concluded that the MRC-MMSE achieves the full antenna array gain at low input SNR and an array gain of $M-K$ for high input SNR.  Lastly, the same matrix inversion used for the UL was shown to also be applicable to the DL after complex conjugation, which drastically reduces the compulational load of the combined TDD processing.


\end{document}